\documentclass[sigconf]{acmart}
\usepackage{graphicx}
\usepackage{algorithm}
\usepackage{rotating}
\usepackage{balance}

\usepackage{tabularx}
\usepackage{subfigure}
\usepackage{bm}
\usepackage{multirow}
\usepackage{enumitem}

\AtBeginDocument{%
  }

\copyrightyear{2025}
\acmYear{2025}
\setcopyright{acmlicensed}
\acmConference[WWW '25]{Proceedings of the ACM on Web Conference}{28 April--2 May, 2025}{Sydney, Australia}
\acmBooktitle{Proceedings of the ACM on Web Conference (WWW '25), 28 April--2 May, 2025, Sydney, Australia}

\begin{document}

\title{Long-Term Interest Clock: Fine-Grained Time Perception in Streaming Recommendation System}

\author{Yongchun Zhu$^{\dag}$}
\affiliation{%
  \institution{ByteDance}
  \city{Beijing}
  \country{China}}
\email{zhuyc0204@gmail.com}

\author{Guanyu Jiang$^{\dag}$}
\affiliation{%
  \institution{ByteDance}
  \city{Beijing}
  \country{China}}
\email{jiangguanyu@bytedance.com}

\author{Jingwu Chen$*$}
\affiliation{%
  \institution{ByteDance}
  \city{Beijing}
  \country{China}}
\email{chenjingwu@bytedance.com}

\author{Feng Zhang}
\affiliation{%
  \institution{ByteDance}
  \city{Shanghai}
  \country{China}}
\email{feng.zhang@bytedance.com}

\author{Xiao Yang}
\affiliation{%
  \institution{ByteDance}
  \city{Beijing}
  \country{China}}
\email{wuqi.shaw@bytedance.com}

\author{Zuotao Liu}
\affiliation{%
  \institution{ByteDance}
  \city{Shanghai}
  \country{China}}
\email{michael.liu@bytedance.com}

\thanks{$\dag$ Yongchun Zhu and Guanyu Jiang have equal contributions.}
\thanks{$*$ Jingwu Chen is the corresponding author.}

\renewcommand{\shortauthors}{Yongchun Zhu et al.}

\begin{abstract}
User interests manifest a dynamic pattern within the course of a day, e.g., a user usually favors soft music at 8 a.m. but may turn to ambient music at 10 p.m. To model dynamic interests in a day, hour embedding is widely used in traditional daily-trained industrial recommendation systems. However, its discreteness can cause periodical online patterns and instability in recent streaming recommendation systems. Recently, Interest Clock has achieved remarkable performance in streaming recommendation systems. Nevertheless, it models users' dynamic interests in a \textit{coarse-grained} manner, merely encoding users' \textit{discrete} interests of 24 hours from \textit{short-term} behaviors. In this paper, we propose a fine-grained method for perceiving time information for streaming recommendation systems, named \textbf{Long-term Interest Clock} (\textbf{LIC}). The key idea of LIC is adaptively calculating current user interests by taking into consideration the relevance of long-term behaviors around current time (e.g., 8 a.m.) given a candidate item.
LIC consists of two modules: (1) \textbf{Clock-GSU} retrieves a sub-sequence by searching through long-term behaviors, using query information from a candidate item and current time, (2) \textbf{Clock-ESU} employs a time-gap-aware attention mechanism to aggregate sub-sequence with the candidate item. With Clock-GSU and Clock-ESU, LIC is capable of capturing users' dynamic fine-grained interests from long-term behaviors.
We conduct online A/B tests, obtaining +0.122\% improvements on user active days. Besides, the extended offline experiments show improvements as well. Long-term Interest Clock has been integrated into Douyin Music App's recommendation system.

\end{abstract}




\maketitle

\vspace{-0.3cm}
\section{Introduction}\label{sec:1}
Time exerts a remarkable and considerable influence on users' interests during the course of a single day. In the early morning hours, users are prone to being inclined towards content that fosters a sense of calm. Specifically, they may exhibit a preference for listening to soft, soothing music. Towards the late hours of the night, when the body and mind are nearing a state of rest, users are likely to have a predilection for more tranquil and meditative content, such as ambient music. Thus, it is crucial for recommendation systems~\cite{zhang2023modeling,zhu2022personalized,zhu2024interest}, which aims to provide satisfying contents to users according to their current interests, to capture users' dynamic interests in a day.


As a typical common solution to enable models to perceive time information, the hour embedding method~\cite{ping2021user,li2022automatically}, which transforms the hour within a day into hour embeddings, has been widely adopted by the early industrial recommendation systems. These systems typically adopt a daily-trained framework, gathering all samples from a single day and then randomly shuffling them for the training process. In recent years, a growing number of companies have deployed real-time streaming recommendation systems, which presents a new challenge to time perception. Within the streaming framework, at any given moment, all training samples possess identical time characteristics. Additionally, recommendation systems possess the capability to produce tens of millions of samples on an hourly basis. This phenomenon consequently causes the recommendation model to merely adapt to the current time features and discard the information acquired during other time periods. This discreteness of the hour embedding methods can  lead to the emergence of periodic online patterns and bring about instability~\cite{zhu2024interest}, which shows unsatisfying performance in real-time streaming recommendation systems.

\begin{figure*}[t]
	\centering
	\begin{minipage}[b]{0.9\linewidth}
		\centering
		\includegraphics[width=1.\linewidth]{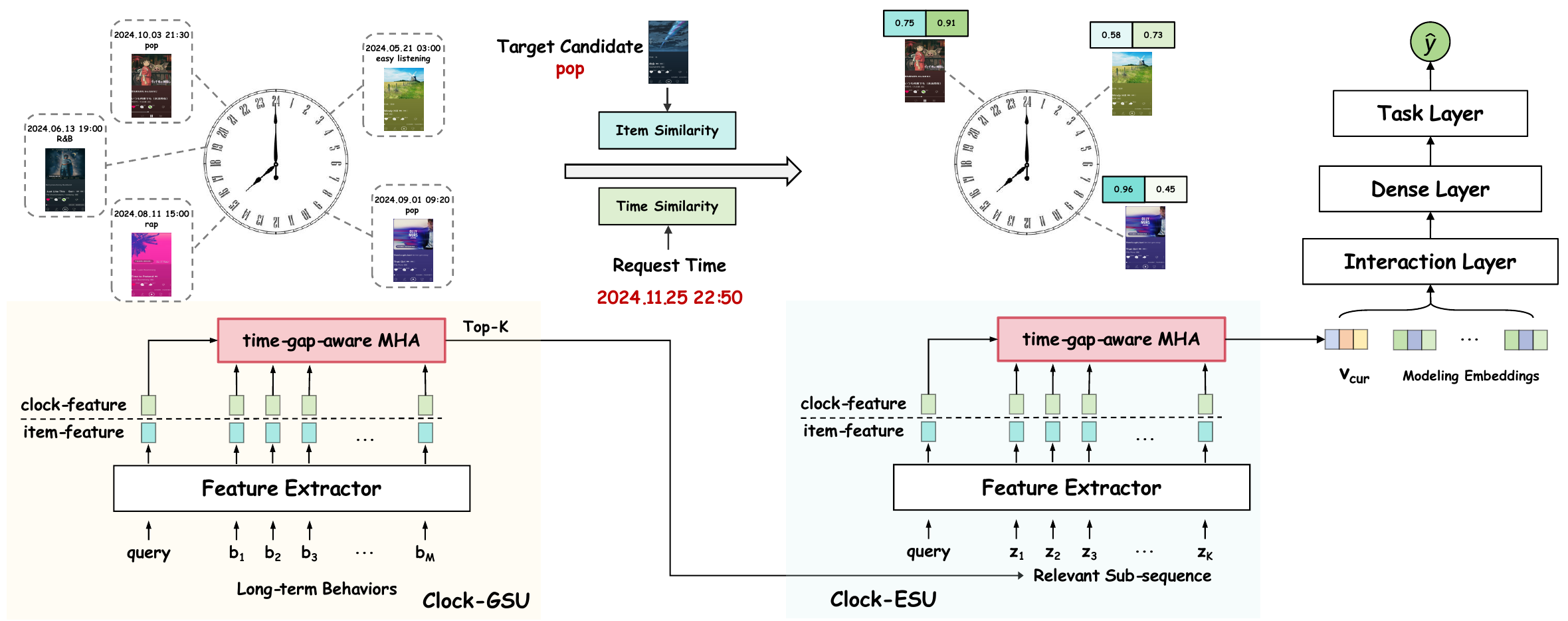}
	\end{minipage}
        \vspace{-0.3cm}
	\caption{Overall architecture of the Long-term Interest Clock.}\label{fig:model}
        \vspace{-0.3cm}
\end{figure*}

Recently, Douyin Group proposed an effective and universal method named Interest Clock~\cite{zhu2024interest} for time perception in streaming recommendation systems. Interest Clock encodes personalized user interests of 24 hours into a clock, which achieves remarkable performance. However, we have observed three drawbacks of this method in our iteration:
\begin{itemize}[leftmargin=1em]
    \item \textbf{Discreteness of interests}: It calculates the user's past interests grouped by hour, so the time-aware feature at 7:59 and 8:01 could be different. Though it utilizes empirical Gaussian distribution to smooth and aggregates the discrete interest embedding, the aggregated embedding is mainly based on adjacent two hours. Thus, if a user has no interaction around the current hour in the past, the time-aware feature will be absent.
    \item \textbf{Coarse-grained interests}: It transforms the user's past behaviors into category-level interests, incorporating only the top three preferences in terms of genre, mood, and language. In the event that diverse users predominantly consume content with the same genre/mood/language around the current hours, they would possess identical time-aware features.
    \item \textbf{Short-term interests}: It calculates discrete coarse-grained time-aware features based on behaviors within the past 30 days, which merely encompasses short-term interests.
\end{itemize}

In this paper, we propose a fine-grained method named \textbf{Long-term Interest Clock} (\textbf{LIC}) to perceive time information in streaming recommendation systems, which adaptively calculates the embedding of current user interests by taking into consideration the relevance of long-term behaviors around current time (e.g., 8 a.m.) given a candidate item. LIC consists of two modules: Clock-based General Search Unit (\textbf{Clock-GSU}) and Clock-based Exact Search Unit (\textbf{Clock-ESU}). Clock-GSU extracts a sub-sequence from long-term behaviors, and the sub-sequence is relevant to a specific candidate item and is centered around the current time. Clock-ESU uses a time-gap-aware attention mechanism, which calculates the relevant score according to the time gap and item similarity and generates the embedding of the user's current interests. Thus, the current interest embedding produced by LIC encompasses the user's \textit{fine-grained} (item-level) \textit{long-term} interests, which demonstrates a \textit{smoothing} effect within a day. The main contributions of our work are summarized in three folds:
\begin{itemize}[leftmargin=1em]
    \item To enable streaming recommendation systems to perceive time information, we propose a novel method named Long-term Interest Clock. To the best of our knowledge, we are the first to tackle the time perception problem using long-term behaviors in real-time streaming recommendation systems.
    \item We conduct online experiments, obtaining +0.122\% improvements on user active days. In addition, offline experiments also demonstrate its effectiveness.
    \item Long-term Interest Clock has been widely deployed in online recommendation systems of Douyin Music App, indicating its superior effectiveness and universality.
\end{itemize}

\vspace{-0.3cm}\section{Related Work}
In this section, we will introduce the related work on Time Perception and Long-term Sequence Modeling.

\textbf{Time Perception}. Hour embedding~\cite{ping2021user,li2022automatically} is a widely adopted method in the industry, which encodes the hour of a day into hour embeddings. However, the hour embedding methods convert time into discrete embeddings, which is ineffective in modern real-time streaming recommendation systems. For takeaway recommendation, \cite{zhang2023modeling} divided a day into four periods, including morning, noon, night, and last night, and used different graph models for different periods, which is difficult to deploy in other scenarios. For streaming recommendation systems, Interest Clock~\cite{zhu2024interest}  encodes personalized user interests of 24 hours into a clock, but it only models short-term discrete interests in a coarse-grained manner.

\textbf{Long-term Sequence Modeling}. Most existing methods~\cite{pi2020search,chang2023twin,feng2024context,sun2024market,si2024twin} adopt a two-stage search framework, which is composed of a General Search Unit and an Exact Search Unit. They mainly consider the similarity between the long sequence and the candidate item. Some of them~\cite{chang2023twin,feng2024context,si2024twin} also take into account the context similarity. However, these methods use hour embedding to represent time information, which is ineffective in streaming recommendation systems. To the best of our knowledge, we are the first to tackle the time perception problem from the perspective of long-term interests in real-time streaming recommendation systems.

\section{Long-term Interest Clock}

In this section, we provide a detailed introduction to the proposed Long-term Interest Clock. In Section~\ref{sec:3.1}, we briefly explain the problem definition. In Section~\ref{sec:3.2}, we introduce the details of Clock-GSU. In Section~\ref{sec:3.3}, we introduce the details of Clock-ESU.

\subsection{Problem Statement}\label{sec:3.1}
In recommendation systems, each sample contains the input raw features and a label $y \in \{0,1\}$, and these features are converted into vectors, named feature embeddings, denoted as $\{\bm{v}_1, \cdots, \bm{v}_N\}$, where $N$ indicates the number of raw features. The prediction of a recommendation model $f(\cdot)$ with the embeddings as inputs is formulated as $\hat{y} = f([\bm{v}_1, \cdots, \bm{v}_N])$. The cross-entropy loss is often used as the optimization target for binary classification:
\begin{equation}
    \mathcal{L} = -y \log \hat{y} - (1-y) \log (1 - \hat{y}).\label{eq:loss}
\end{equation}

In this paper, we focus on extracting the representations of the user's current interests $\bm{v}_{cur}$ according to current time $t^{cur}$, the candidate item and long-term behaviors. The candidate item contains lots of item features, and the concentrated embedding is utilized as a query denoted as $\bm{q}  \in \mathbb{R}^{H}$. Each behavior contains an item and its features, and the behavior embedding is denoted as $\bm{b}  \in \mathbb{R}^{L}$, where $L$ denotes the dimension of $\bm{b}$. Thus, long-term behaviors is denoted as $\{\bm{b}_1, \cdots, \bm{b}_M\}$, and $M$ indicates the length of long sequence. The time at which the $m$-th behavior occurs is defined as $t^{b_m}$.

\begin{table*}[htbp]
\centering
\setlength\tabcolsep{7pt}
\caption{Online A/B testing results of a ranking task. The results indicate the relative improvement with our Long-term Interest Clock over the baseline. The square brackets represent the 95\% confidence intervals for online metrics. Statistically significant improvement is marked with bold font in the table.}
\begin{tabular}{lccccc}
\toprule
\multirow{2}{*}{}                    & \multicolumn{1}{c}{Main Metrics}                    & \multicolumn{4}{c}{Constraint Metrics}                                                                    \\
\cmidrule(r){3-6}
                                     & Active Day                            & Like                     & Finish                   & DisLike                 & Play                     \\
\midrule
\multirow{2}{*}{\textbf{Long-term Clock}}             & \textbf{0.122\%}                 & \textbf{0.101\%}         & \textbf{0.093\%}         & \textbf{-0.683\%}         & \textbf{0.095\%}         \\
                                     & {[}-0.052\%, +0.052\%{]} &  {[}-0.083\%, +0.083\%{]} & {[}-0.065\%, +0.066\%{]} & {[}-0.586\%, +0.586\%{]} & {[}-0.079\%, +0.079\%{]}\\
\bottomrule
\end{tabular}\label{tab:online}
\end{table*}

\begin{table}[htbp]
\centering
\caption{Offline results (AUC, UAUC and RelaImpr) on the industrial datasets DouyinMusic-20B.}
\vspace{-0.2cm}
\begin{tabular}{lcccc}
\toprule
               & AUC  & RelaImpr  & UAUC  & RelaImpr   \\
\midrule
Base Model    & 0.6643 & - &  0.6023 & - \\
Hour Embedding & 0.6631 & -0.18\% &  0.6007 & -0.27\% \\
Naive Clock    & 0.6666 & +0.35\% & 0.6015 & -0.13\% \\
Adaptive Clock & 0.6662 & +0.29\% &  0.5859 & -2.72\% \\
Gaussian Clock & 0.6695 & +0.78\% & 0.6069 & +0.76\%\\
Long-term Clock & \textbf{0.6720} & \textbf{+1.16\%} & \textbf{0.6113} & \textbf{+1.49\%} \\
\bottomrule
\end{tabular}\label{tab:offline}
\end{table}

\subsection{Clock-Based General Search Unit}\label{sec:3.2}
Clock-GSU aims to extract a sub-sequence from long-term behaviors, and the sub-sequence is relevant to the candidate item $q$ and is around the current time $t^{cur}$. Firstly, we define a function $g(\cdot)$ to extract hours, minutes, and seconds, e.g., $g(2024.12.10\ \ 13:30:00) = 13:30:00$. Then, the relative time gap between two behavior $\bm{b}^{n}$ and $\bm{b}^{m}$ is indicated as $\Delta(t^{b_n}, t^{b_m})$, e.g., $\Delta(23:00:00, 1:00:00) = 120(minutes)$ and $\Delta(11:00:00, 17:00:00) = 360(minutes)$.

The relevant score $\alpha \in \mathbb{R}$ between the candidate item and a historical behavior is calculated as:
\begin{equation}\label{eq:attention}
    \alpha(\bm{b}_m, \bm{q}) = \underbrace{\frac{(W_b \times \bm{b}_m)\odot(W_q \times \bm{q})^T}{\sqrt{d}}}_{\textbf{item similarity}} + \underbrace{s(\Delta(t^{b_m}, t^{cur}))}_{\textbf{time similarity}},
\end{equation}
where $W_b \in \mathbb{R}^{d \times L}$ and $W_q \in \mathbb{R}^{d \times H}$ are learnable parameters, and $d$ indicates the dimension of latent vectors. $s(\cdot)$ is a two-layer neural networks with $[\Delta, \sqrt{\Delta}, \Delta^2, \log(\Delta + 1)]$ as input. With the relevant scores, Clock-GSU can proceed with the top-K search over ten thousand length of user behaviors. The top-K sub-sequence is denoted as $\{\bm{z}_1, \cdots, \bm{z}_K\}$, where $K=100$ in LIC.

Note that, different from our proposed relative time gap, most existing context-aware methods~\cite{pi2020search,chang2023twin,feng2024context,si2024twin} use absolute time gap as input, which cannot model user's dynamic interests within a day in streaming recommendation systems. To further speed up the top-K search, the results of $W_b \times \bm{b}_m$ and $W_q \times \bm{q}$ are pre-computed and stored in online parameter servers (PS). This allows for convenient direct access and retrieval of the results, facilitating a rapid computation of the relevant score. The parameters of $s(\cdot)$ with dimensions $[8, 1]$ are also stored in online PS. The structure of $s(\cdot)$ is simple, so the time complexity is substantially lower than $(W_b \times \bm{b}_m)\odot(W_q \times \bm{q})^T$.

\subsection{Clock-Based Exact Search Unit}\label{sec:3.3}
With Clock-GSU, the sub-sequence of top-K items $\{\bm{z}_1, \cdots, \bm{z}_K\}$, which is relevant to the candidate item around current time, is retrieved from long-term behaviors. Clock-ESU aims to extract the representations of the user's current interests, denoted as $\bm{v}_{cur}$.

Inspired by multi-head attention in long-term sequential methods~\cite{pi2020search,chang2023twin,si2024twin}, we utilize a multi-head time-gap-aware attention mechanism to calculate the relevant score. Let $Z \in \mathbb{R}^{K \times L}$ indicate the matrix of top-K behaviors $\{\bm{z}_1, \cdots, \bm{z}_K\}$. To further enhance the time perception capability of Clock-ESU, we concentrate $[\Delta, \sqrt{\Delta}, \Delta^2, \log(\Delta + 1)]$ into representations of top-K behaviors $\bm{z}$. Thus, the representation of one head $\bm{r}_i \in \mathbb{d}$ is denoted as:
\begin{equation}
    \bm{r}_i = \textit{Softmax}(\bm{\alpha}_i)^T Z W_{vi},
\end{equation}
where $W_{vi} \in \mathbb{R}^{L \times d}$ is the learnable parameter matrix of the $i$-th head, and $\bm{\alpha}_i$ indicates the top-K relevant scores of the $i$-th head, computed with Equation~(\ref{eq:attention}). Note that each head has different parameter matrices $W_b$ and $W_q$. In practice, the number of heads is four, and the representation of the user's current interests is formulated as:
\begin{equation}
    \bm{v}_{cur} = h([\bm{r}_1, \cdots, \bm{r}_4]),
\end{equation}
where $h(\cdot)$ is a two layer deep network. The user's current interests $\bm{v}_{cur}$ are fed into $f(\cdot)$ for final predictions.

Note that both Clock-GSU and Clock-ESU incorporate the aspect of time similarity. As a result, Clock-GSU is capable of retrieving relevant items in proximity to the current time, while Clock-ESU can adaptively model the relevance of the retrieved top-K behaviors based on both item similarity and time similarity. With Clock-GSU and Clock-ESU, the proposed Long-term Interest Clock can adaptively model users' current interests. In addition, both the two modules utilize the multi-head attention mechanism. In Clock-GSU, $(W_b \times \bm{b}_m)$ and $(W_q \times \bm{q})$ of each head are pre-computed and stored in online PS. In practical computation, we directly concatenate the pre-computed embeddings of each head to calculate item similarity.

\section{Experiments}


In this section, we conduct extensive offline and online experiments to demonstrate the effectiveness of the propose method.

\textbf{Datasets.} We evaluate the Long-term Interest Clock with baselines on a large-scale industrial recommendation dataset. Note that Long-term Interest Clock requires a substantial amount of data with long-term behaviors (spanning one year) from streaming recommendation systems. However, currently, there is no public dataset that is suitable for this purpose. Thus, we only adopt the industrial dataset DouyinMusic-20B from our system.

\textit{DouyinMusic-20B}: Douyin offers a music recommendation service, which has over 10 million daily active users. ~\citet{zhu2024interest} collect from the impression logs and get one dataset. The dataset contains more than 20 billion samples, denoted as \textit{DouyinMusic-20B}. Each sample in the industrial dataset includes over a hundred features. These features consist of non-ID meta features such as gender, age, genre, mood, scene, etc., as well as ID-based personalized features like user ID, item ID, artist ID, and interacted ID sequence, which can effectively represent real-world scenarios. The DouyinMusic-20B dataset contains samples from Douyin Music across the time span of 8 weeks from August to September 2023. Following~\cite{zhu2024interest}, we use `Finish' as the label and take the first 6 weeks as the training set, the following 1 week as the validation set, and the remaining 1 week as the test set. 

\textbf{Online A/B Testing.} To verify the real benefits Long-term Interest Clock brings to our system, we conducted online A/B testing experiments for more than one month for the ranking task in Douyin Music App. We evaluate model performance based on one main metric, Active Days. We also take additional metrics, which evaluate user engagement, including Like, Finish, DisLike, and Play, which are usually used as constraint metrics~\cite{zhu2024interest}. We apply the proposed Interest Clock on a DCN-V2-based multi-task model~\cite{wang2021dcn} which is deployed in the online ranking tasks. The online A/B results are shown in Table~\ref{tab:online}. For the main metrics Active Days, the proposed Interest Clock achieves a large improvement of +0.122\% for all users with statistical significance, which is remarkable given the fact that the average Active Days improvement from production algorithms is around 0.05\%.

\textbf{Offline Results.} We utilize AUC, UAUC, and the relative improvements (RelaImpr) as offline metrics. We compare the proposed LIC with (1) Base Model is a DCN-V2-based model~\cite{wang2021dcn} without any time modeling method, (2) Hour Embedding, (3) Naive Clock, (4) Adaptive Clock, and (5) Gaussian Clock. It should be noted that all the baseline models employ the same base model, and Naive/Adaptive/Gaussian Clock methods are the same as ~\citet{zhu2024interest}. The experimental results on the industrial dataset are shown in Table~\ref{tab:offline}. The results further reveal several insightful observations. Hour Embedding method proves ineffective in streaming data. UAUC of Adaptive Clock is inferior to that of the baseline, potentially because the adaptive weights of time information are challenging to learn in streaming recommendation systems. Long-term Interest Clock could outperform the best baseline Gaussian Interest Clock significantly, which demonstrates empirical Gaussian weights are effective.

\section{Conclusion}
In this paper, with the aim of endowing streaming recommendation systems with the ability to perceive time alterations, we propose an effective method named Long-term Interest Clock (LIC).
LIC consists of a Clock-GSU and a Clock-ESU. Clock-GSU retrieves a sub-sequence from long-term behaviors, which is around the current time and relevant to a candidate item.  Clock-ESU employs a time-gap-aware attention mechanism to extract a representation of the user's current interests from the sub-sequence. 
We demonstrated the superior performance of the proposed Long-term Interest Clock in both offline and online experiments, which demonstrates its effectiveness. 
Moreover, Long-term Interest Clock has been deployed on ranking tasks in multiple applications of Douyin Group.

\bibliographystyle{ACM-Reference-Format}
\bibliography{sigconf}

\appendix



\end{document}